\documentclass[journal]{IEEEtran}
\usepackage{amsmath}
\usepackage{array}
\usepackage{multirow}
\usepackage{booktabs}
\usepackage{color}
\usepackage{amssymb}
\usepackage{diagbox}
\usepackage{graphicx}
\usepackage{subfigure}
\usepackage{cite}
\usepackage{float}

\usepackage[colorlinks,linkcolor=red,citecolor=green,urlcolor=blue]{hyperref}

\makeatletter
\def\ps@IEEEtitlepagestyle{%
  \def\@oddfoot{\mycopyrightnotice}%
  \def\@evenfoot{}%
}
\def\mycopyrightnotice{%
  {\hfill \scriptsize {This work has been submitted to the IEEE for possible publication.
  Copyright may be transferred without notice, after which this version may no longer be accessible.}\hfill}

}
\makeatother

\begin{document}
\title{Two-stream Hierarchical Similarity Reasoning for Image-text Matching}

\author{
        Ran~Chen,~\IEEEmembership{}
        Hanli~Wang,~\IEEEmembership{Senior Member,~IEEE,}
        Lei~Wang,~\IEEEmembership{}
        Sam Kwong,~\IEEEmembership{Fellow,~IEEE}

        \thanks{\emph{Corresponding author: Hanli Wang.}}

        \thanks{R.~Chen and H.~Wang are with the Department of Computer Science \& Technology, Key Laboratory of Embedded System and Service Computing (Ministry of Education), Tongji University, Shanghai 200092, P. R. China, and with Frontiers Science Center for Intelligent Autonomous Systems, Shanghai 201210, P. R. China (e-mail: 2110142@tongji.edu.cn, hanliwang@tongji.edu.cn).}

        \thanks{Lei Wang is with DeepBlue Academy of Sciences, Shanghai 200336, P. R. China (e-mail: wangl@deepblueai.com).}

        \thanks{Sam Kwong is with the Department of Computer Science, City University of Hong Kong, Hong Kong, P. R. China (e-mail: cssamk@cityu.edu.hk).}}

\markboth{}%
{Shell \MakeLowercase{}}

\maketitle

\begin{abstract}
Reasoning-based approaches have demonstrated their powerful ability for the task of image-text matching. In this work, two issues are addressed for image-text matching. First, for reasoning processing, conventional approaches have no ability to find and use multi-level hierarchical similarity information. To solve this problem, a hierarchical similarity reasoning module is proposed to automatically extract context information, which is then co-exploited with local interaction information for efficient reasoning. Second, previous approaches only consider learning single-stream similarity alignment (\emph{i.e.}, image-to-text level or text-to-image level), which is inadequate to fully use similarity information for image-text matching. To address this issue, a two-stream architecture is developed to decompose image-text matching into image-to-text level and text-to-image level similarity computation. These two issues are investigated by a unifying framework that is trained in an end-to-end manner, namely two-stream hierarchical similarity reasoning network. The extensive experiments performed on the two benchmark datasets of MSCOCO and Flickr30K show the superiority of the proposed approach as compared to existing state-of-the-art methods.
\end{abstract}


\section{Introduction}

The image-text matching task aims to measure the visual-semantic similarity between image and text, which has a lot of potential applications such as cross-modal retrieval~\cite{01wacv/csgm, 01tip/MABAN, 03tip/Hashing}, image captioning~\cite{01cvpr/bottom-up, 01tmm/captionnet, 02tip/Topic-Oriented}, text-to-image synthesis~\cite{02cvpr/AttnGAN}, and multi-modal neural machine translation~\cite{2016arxiv/neural}. The traditional methods~\cite{01ijcai/caran,01nc/cca,01ijcv/Multi-View} constructed common space to maximize the correlation of cross-media information. However, these methods are limited by the hand-crafted features, which require strong prior knowledge. To address this issue, learning based methods have been applied to extract features directly from different modalities. In recent years, research efforts have been devoted to learn the representations of both image and text, and match different modalities based on the learned representations. To this aim, several works~\cite{03cvpr/Structure-Preserving,2018arxiv/vsepp} mapped the entire image and the full text into a common space, followed by computing the cosine similarity in the common space. However, these approaches have an insufficiently discriminative ability due to the lack of local interaction between sentence words and image regions. What's worse, although several techniques~\cite{01eccv/scan,01iccv/camp} have been proposed to extract local interaction, it is difficult to match more complicated image-text pairs by simply grasping these local semantic concepts but overlooking higher-level information.

\begin{figure*}[t!]
     \centering
     \includegraphics[width=0.85\linewidth]{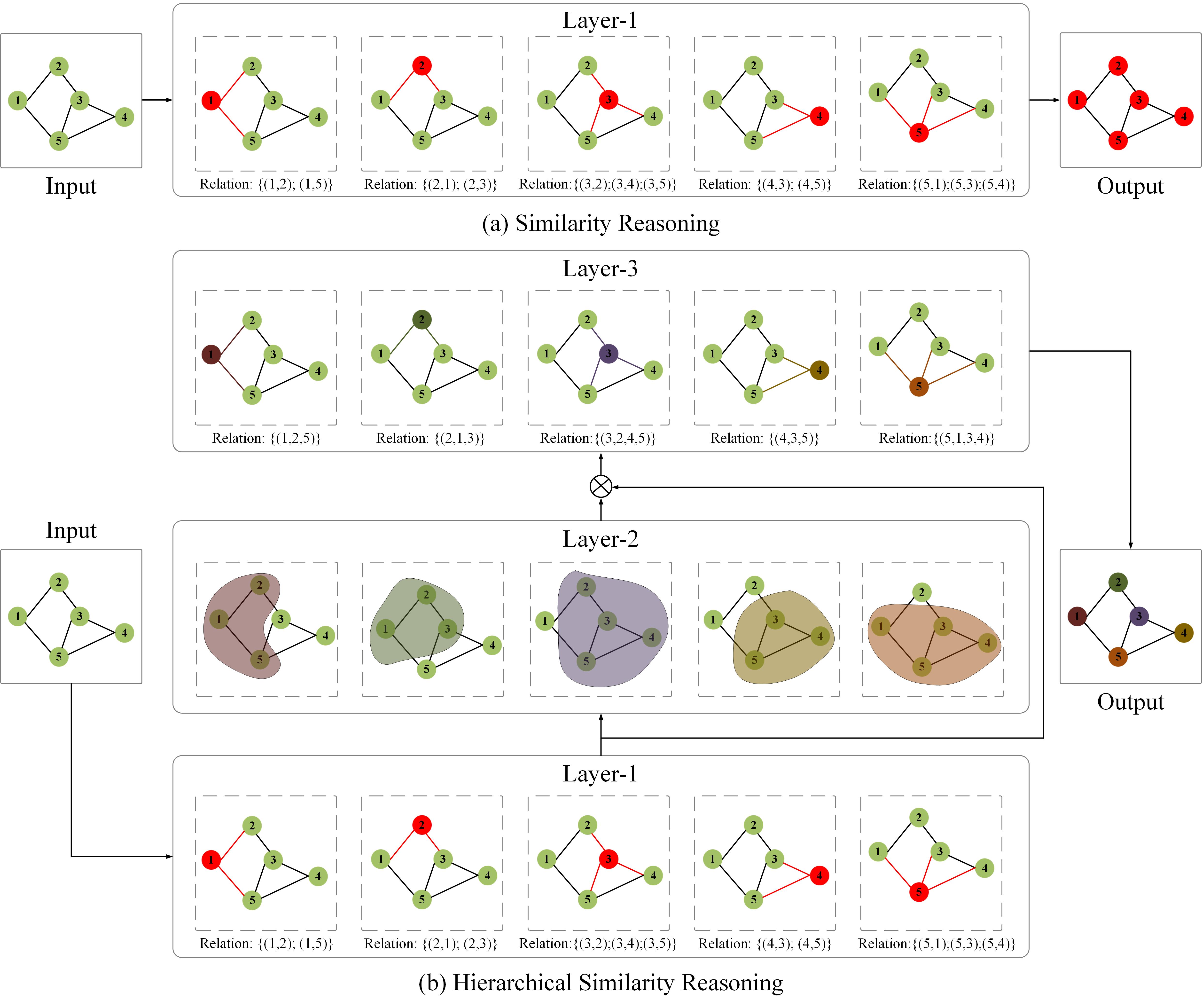}
     \caption{Illustration of the difference between similarity reasoning~\cite{01aaai/sgraf} and the proposed hierarchical similarity reasoning. Both of these two reasoning strategies start with propagating nearest neighbor information, then (a) the similarity reasoning strategy forms global similarity representations based on the propagated nearest neighbor information directly and ignores the multi-level characteristic of interaction structure, while (b) the proposed hierarchical similarity reasoning strategy automatically finds higher-level information based on the propagated nearest neighbor signals and investigates multi-level signals to form the global information gradually.}
     \label{principle}
 \end{figure*}

Recently, higher-level information can be extracted through reasoning-based~\cite{02iccv/vsrn,01aaai/sgraf} or attention-based~\cite{02ijcai/mvsa} methods. In general, reasoning-based methods firstly adopt different semantics as vertexes to construct an undirected graph. Afterwards, each vertex communicates with its nearest neighbor vertexes. Finally, global representation can be obtained by aggregating all of the vertexes. Specifically, Li~\textit{et al}~\cite{02iccv/vsrn} used salient regions as vertexes to reason region relationships for enhancing the original visual features. Unfortunately, limited fine-grained matching is arisen because of unconsidered region-word local interactions. Diao~\textit{et al}~\cite{01aaai/sgraf} exploited region-word local similarities as vertexes so that higher-level similarity information was obtained via reasoning these local similarities. However, as shown in Fig.~\ref{principle}(a), the conventional reasoning mechanism tries to realize global representation simply through information communication between directly connected vertexes, thus each vertex ignores to excavate latent information with the non-directly connected vertexes, for instance context information. About attention-based methods, a relation-wise dual attention network~(RDAN) was devised in~\cite{02ijcai/mvsa} which not only extracted salient objects and key words, but also explored latent relations between objects and words. However, there are two limitations. First, RDAN excavates higher-level information directly based on the cross-attentive map which is misaligned with local similarity, hence the problem of heterogeneity gap cannot be solved and the multilevel information is not accurate enough. Second, due to the lack of reasoning mechanism, RDAN is hard to effectively calculate global similarity.

\begin{figure}[htbp]
    \centering
    \includegraphics[width=0.85\linewidth]{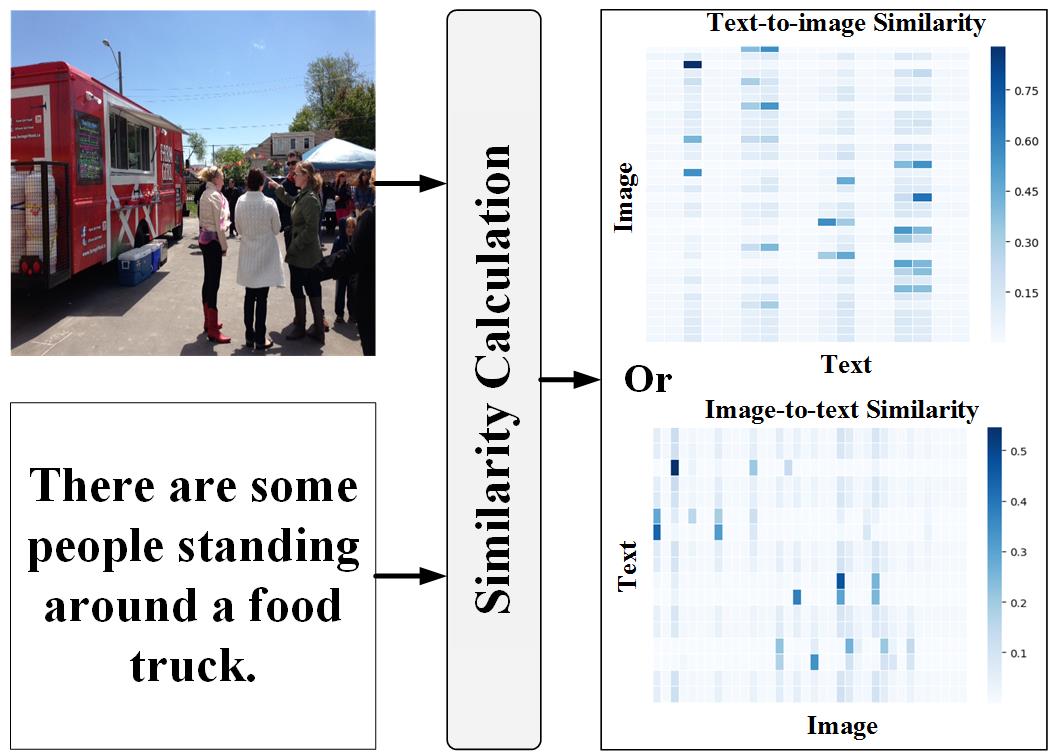}
    \caption{Illustration of similarity gulf between image-to-text and text-to-image calculations.}
    \label{two-branch}
\end{figure}

In general, image-text interaction should be of hierarchy~\cite{15cvpr/mcnn,16aaai/text,02ijcai/mvsa}, progressively from local interaction to context-awareness and then to global representation. Concretely, local interaction usually attends to sentence words about image regions; context-awareness attempts to capture several latent information in a local receptive field, such as relative position information in a local space; global representation is more robust by aggregating multi-level information and allowing information communication. However, it is still hard to automatically find and fully utilize hierarchical information without the guidance of external auxiliary knowledge, such as knowledge mapping~\cite{04ijcai/Knowledge} and syntactic analysis~\cite{01pami/cmpc}. Moreover, as shown in Fig.~\ref{two-branch}, the similarities between image-to-text and text-to-image calculations are different and complementary. Most existing works such as~\cite{02iccv/vsrn,01aaai/sgraf,02ijcai/mvsa} computed similarity only by adopting image features as cues to attend on texts (image-to-text similarity) or vice versa (text-to-image similarity), where the similarity of image-to-text and text-to-image is not considered in a unified network.

In order to explore hierarchical information sufficiently for the image-text matching task, a two-stream hierarchical similarity reasoning~(TSHSR) network is designed in this work. In detail, a hierarchical similarity reasoning~(HSR) module is proposed to fully utilize hierarchical information to progressively form a more robust similarity representation. As shown in Fig.~\ref{principle}(b), there are three layers with different functions in HSR. In the first layer, HSR starts with computing region-word local similarities, and takes these similarities as vertexes, then each vertex exchanges information with its nearest neighbors. In the second layer, to explore context information, an assembling operation is adopted to comprehensively investigate each vertex and its nearest neighbors as a whole, as well as fuse lower-level information from the first layer via the operation of element-wise multiplication. Consequently, each vertex collects the information from its nearest neighbors while attending to context information as well. In the third layer, the vertexes own more abundant information which boosts more proper reasoning to satisfy sophisticated matching patterns. What's more, both image-to-text similarity and text-to-image similarity are utilized to enrich and supplement neglected global representations, so that global representations can have a stronger representative ability. Moreover, TSHSR is an end-to-end trainable framework.

There are three major contributions of this work. First, a HSR module is proposed for hierarchical similarity reasoning. Different from other reasoning architectures which suffer from the problem of insufficient hierarchical information, HSR enables to automatically recognize multi-level information, which is exploited in an efficient integration manner to achieve more robust global representations. Second, both the image-to-text similarity and text-to-image similarity are utilized to enhance image-text mutual retrieval, since these two similarities are complemented to each other. Third, a TSHSR framework is designed to fully explore the usage of the aforementioned two similarities with a hierarchical similarity reasoning strategy, and an efficient and effective way is devised for TSHSR to optimize its modules in an end-to-end manner. The rest of this paper is organized below. Section~\ref{sect:rewk} reviews the related works. Section~\ref{sect:prop} describes the algorithm of the proposed TSHSR, and Section~\ref{sect:exp} presents experiments to validate the effectiveness of TSHSR. Finally, Section~\ref{sect:cln} concludes this work.

\section{Related Work}
\label{sect:rewk}

In this section, feature encoding, image-text matching and graph reasoning are reviewed.

\subsection{Feature Encoding}
\label{sect:rewk:fe}

Feature encoding plays an important role for image-text matching. For text encoding, conventional methods~\cite{02nips/DeViSE,04cvpr/Fisher} tried to utilize SkipGram~\cite{01iclrw/Efficient} or Fisher vectors~\cite{05cvpr/Fisher} to encode textual features. However, these methods were unable to capture sequential information. To address this issue, Kiros~\textit{et al}~\cite{01icml/Unifying} adopted a GRU~\cite{01nipsw/Empirical} instead of SkipGram as text encoder. Regarding image encoding, it can be roughly divided
into two classes: coarse-grained representation and fine-grained representation. The efforts of coarse-grained representation try to encode more robust global image representations without resorting to salient features, \emph{e.g.}, Liu~\textit{et al}~\cite{03iccv/rrfn} proposed a recurrent residual network that could refine global embeddings, Song~\textit{et al}~\cite{07cvpr/Polysemous} and Wei~\textit{et al}~\cite{14cvpr/Multi-Modality} both combined global context with locally-guided features employing multi-head self-attention. Besides, some works~\cite{06cvpr/dan,04iccv/sgan} paid attention to gather semantics by exploiting block-based visual attention on feature maps. Differently, the fine-grained representation uses local salient features for more detailed learning, with the motivation inspired by bottom-up attention~\cite{01cvpr/bottom-up}. To this aim, several methods~\cite{01eccv/scan,01iccv/camp,02iccv/vsrn,01wacv/csgm,02aaai/Expressing,01aaai/sgraf} employed object detectors (\emph{e.g.}, faster-RCNN~\cite{03nips/faster-rcnn}) pre-trained on a large-scale dataset to capture region-based features of visual objects. And then, various algorithms are proposed based on the captured region-based features to calculate global representation, \emph{e.g.}, Chen~\textit{et al}~\cite{02aaai/Expressing} exploited a BiGRU to obtain high-level semantic features, Li~\textit{et al}~\cite{02iccv/vsrn} introduced a visual reasoning mechanism to build the relationship between fine-grained visual features. Instead of reasoning on visual features, which has no interaction with textual features, Diao~\textit{et al}~\cite{01aaai/sgraf} concentrated on establishing the relationship between constructed visual-textual local similarities. Inspired by this, local similarity representation is employed in this work to progressively form hierarchical interaction, which is different from using local similarity representation directly in~\cite{01aaai/sgraf}.

\subsection{Image-Text Matching}
\label{sect:rewk:itm}

After feature encoding, it is required to calculate the final similarity across modalities according to a certain alignment mechanism for image-text matching. Existing alignment methods could be classified to two categories: global alignment~\cite{2018arxiv/vsepp,03cvpr/Structure-Preserving,03iccv/rrfn,07cvpr/Polysemous,06cvpr/dan,02iccv/vsrn} and local alignment~\cite{08cvpr/dvsa,01eccv/scan,02ijcai/mvsa,13cvpr/Neighbourhood,09cvpr/imram}. About global alignment, it usually constructs a joint space, and then conducts similarity computation between global image-text representations in this space. Besides, the methods~\cite{02iclr/Order-embeddings,10cvpr/Look} focused on measuring antisymmetric visual-semantic hierarchy by introducing an ordered representation instead of directly computing similarity. As for local alignment, typical methods prefer to compute all region-word pairs and fuse all possible pairs. In general, global-alignment methods have the ability to capture global semantic information while local-alignment methods tend to grasp regional information. By integrating local alignment and global alignment, Diao~\textit{et al}~\cite{01aaai/sgraf} proposed to aggregate region-word local similarities. However, existing methods only consider alignment in the image-to-text level, but ignore the text-to-image level alignment which is also important for cross-media retrieval. In this work, both image-to-text level and text-to-image level alignments are applied to enrich cross-attentive information for mutual retrieval.

\subsection{Graph Reasoning}

It is beneficial to make information communicated between extracted local regions for enhancing global representation. To address the problem of insufficient communication of information, several graph reasoning approaches~\cite{01nips/gcn_Molecular_Fingerprints,01iclr/gcn_semi,03iclr/gated} have been extensively used for cross-media computing, such as image captioning~\cite{11cvpr/Auto-Encoding} and grounding referring expressions~\cite{13cvpr/Neighbourhood}. For image-text matching, Shi~\textit{et al}~\cite{04ijcai/Knowledge} aimed to construct scene concept graph with image scene graphs and co-occurring concept pairs, Li~\textit{et al}~\cite{02iccv/vsrn} introduced graph convolutional network to build up relation between image regions, Wang~\textit{et al}~\cite{01wacv/csgm} adopted textual and visual scene graphs to refine textual and visual features, Wen~\textit{et al}~\cite{01tcsvt/dsran} employed graph attention~\cite{04iclr/gan} on both visual and textual features for semantic relation reasoning. However, these works have no ability to establish more complex matching patterns because hierarchical information is neglected. In this work, a novel graph reasoning mechanism is designed to effectively excavate hierarchical information for cross-media interaction.

\begin{figure*}[t!]
    \centering
    \small
    \includegraphics[width=1\textwidth]{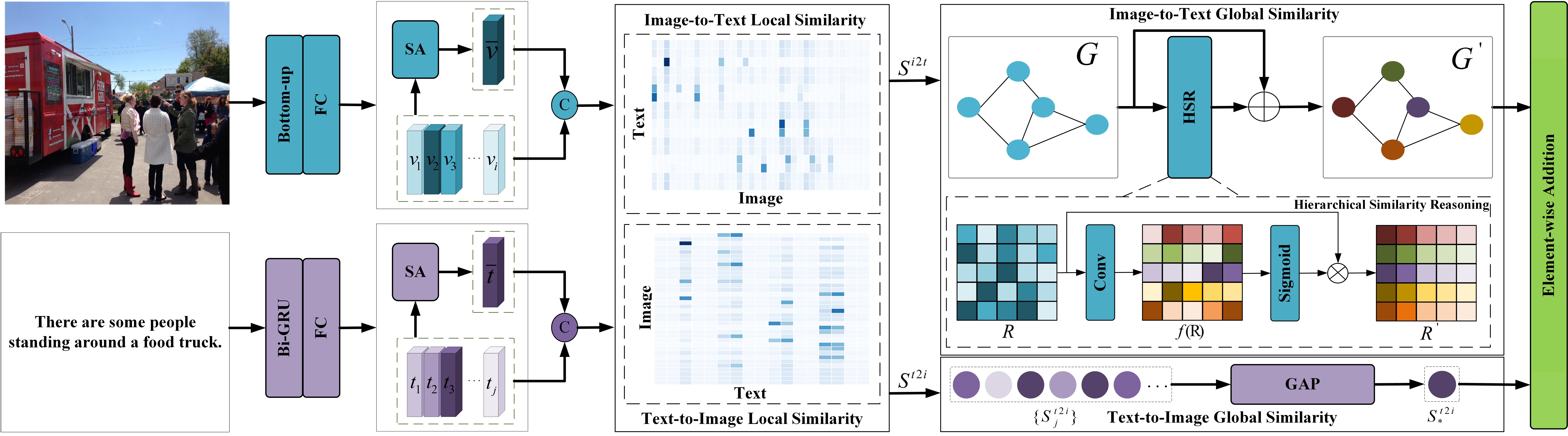}
    \caption{An overview of the proposed TSHSR framework. For image-to-text similarity, after aligning image-text local features, the HSR module is utilized for global image-to-text similarity representation. For text-to-image similarity representation on the other hand, visual and textual local features are aligned firstly, and then global average pooling is used for global text-to-image similarity representation. An element-wise addition operation is applied to summarize these two global representations.}
    \label{network}
\end{figure*}

\section{Proposed Two-stream Hierarchical Similarity Reasoning for Image-text Matching}
\label{sect:prop}

The framework of the proposed TSHSR is illustrated in Fig.~\ref{network}. Technically, for image-to-text similarity representation, aligning image-text local features is firstly started by attending to the sentence word about each image region. Afterwards, the HSR module is used to calculate the global image-to-text similarity. Simultaneously, the text-image local features are firstly aligned via attending to the regions in the sentence with respect to each sentence word, then global average pooling is applied to achieve global text-to-image representation. In the end, these two global representations are summed up to obtain the final similarity representation.

\subsection{Feature Encoding}
\label{sect:prop:fe}

\subsubsection{Image Feature}
\label{sect:prop:fe:img}

For each input image, $K$ region-level visual features~\cite{01cvpr/bottom-up} are extracted firstly, and then a linear transformation is employed to embed them into $d$-dimensional vectors as local region representations $V=\{v_1,\cdots,v_k\}$, with $v_i \in R^d$. Inspired by~\cite{01aaai/sgraf}, the global visual feature $\overline{v} = V \odot \overline{q}_v$ is computed, where $\overline{q}_v=\frac{1}{K}\sum_{i=1}^K v_i$ is calculated to make up for the shortcoming of lacking global embedding, the operator $\odot$ represents element-wise multiplication.

\subsubsection{Text Feature}
\label{sect:prop:fe:txt}

Given a sentence, it is firstly split into $L$ word tokens with tokenization technique followed by previous works~\cite{02iccv/vsrn,01eccv/scan,01aaai/sgraf}, and then the tokens are fed into a BiGRU~\cite{01tsp/birnn} sequentially to obtain the representation $T=\{t_1,\cdots,t_L\}$, with $t_j \in R^d$. Similar to image embedding, global text embedding are also computed by $\overline{t}=T \odot \overline{q}_t$, where $\overline{q}_t=\frac{1}{L}\sum_{j=1}^L t_j$.

\subsection{Local Similarity Representation}
\label{sect:prop:lsr}

Let $S(x,y)$ denote the similarity calculation between $x$ and $y$. Compared with previous works~\cite{02iccv/vsrn,01eccv/scan} that immediately used scalar to calculate cosine or Euclidean distance, Diao~\textit{et al}~\cite{01aaai/sgraf} used a vector to calculate the distance between modalities to preserve more detailed information by
\begin{equation}
	S(x,y)=\frac{{W\left| x-y \right|}^2}{{\left \| x-y \right \|}_2},
\end{equation}
where $W=R^{m \times d}$ is a learnable parameter matrix for obtaining $m$-dimensional similarity vector, while ${\left| \cdot \right|}^2$ and ${\left \| \cdot \right \|}_2$ indicate element-wise square and ${l_2}$-norm, respectively. Therefore, the similarity between $\overline{v}$ and $\overline{t}$ is computed as $s^g=S(\overline{v},\overline{t})$.

When local regions are guided to focus on the attended words, the following similarity representation $s^{i2t}=S({\alpha}_j^v,t_j)$ is obtained, where ${\alpha}_j^v=\sum_{i=1}^K {{\alpha}_{ij}}v_i$ is the attended visual feature with respect to $j$-th word, and ${\alpha}_{ij}$ is the attention weight between region feature $v_i$ and word feature $t_j$.  ${\alpha}_{ij}$ is computed by
\begin{equation} {\alpha}_{ij}=\frac{E({\lambda}\overline{c}_{ij})}{\sum_{i=1}^K{E({\lambda}\overline{c}_{ij})}}, 	\overline{c}_{ij}=\frac{[c_{ij}]_+}{\sqrt{\sum_{j=1}^L[c_{ij}]_+^2}},
\end{equation}
where $\lambda$ is a temperature parameter, $[c_{ij}]_+ = \max(0,\cos(v_i,t_j))$, $\cos(\cdot)$ stands for cosine distance. Similarly, the text-to-image similarity $s^{t2i}=S(v_i, {\beta}_i^t)$ is generated, where ${\beta}_i^t=\sum_{j=1}^L{\beta}_{ij}t_j$ is the attended textual feature with respect to $i$-th region. ${\beta}_{ij}$ is the attention weight between region feature and word feature, and it is computed by
\begin{equation}
{\beta}_{ij}=\frac{E({\lambda}\overline{c}_{ij})}{{\sum_{j=1}^L}E({\lambda}\overline{c}_{ij})},
\overline{c}_{ij}=\frac{[c_{ij}]_+}{\sqrt{\sum_{i=1}^K{[c_{ij}]_+^2}}}.
\end{equation}

\subsection{Global Similarity Representation}
\label{sect:prop:gsr}

After obtaining both the image-to-text and text-to-image  local similarity representations, the next step is to learn a global similarity representation for image-text matching. To achieve this, $s^{i2t}$ and $s^g$ are firstly concatenated to form an initial global feature $S^{i2t}=\{S_1^{i2t},\cdots,S_L^{i2t},S_{L+1}^{i2t}\}$, then the proposed HSR is used for hierarchical similarity reasoning. As shown in Fig.~\ref{network}, a similarity relationship is established between $S_p^{i2t} \in S^{i2t}$ and $S_q^{i2t} \in S^{i2t}$ as
\begin{equation}
	\mathcal{R}(S_p^{i2t}, S_q^{i2t}) = {\varphi}(S_p^{i2t})^T{\phi}(S_q^{i2t}),
\label{eq:affine}
\end{equation}
where ${\varphi}(S_p^{i2t})={W_p}{S_p^{i2t}}$ and ${\phi}(S_q^{i2t})={W_q}{S_q^{t2i}}$ are two embeddings, $W_p$ and $W_q$ are learnable parameters.

Afterwards, a similarity relationship graph $G=(\mathcal{S};\mathcal{R})$ is constructed, where $\mathcal{S}$ is the set of calculated local similarity representations described by $S^{i2t}$ and $\mathcal{R}$ is the relation matrix obtained by calculating the similarity relationship in Eq.~\eqref{eq:affine}. Furthermore, a convolution operation is used to assemble $n$-gram and $n$-term information in a local receptive field for context-awareness. $n$-gram occurs with $n$ successive words, and $n$-term allows for order or semantic alternatives~\cite{16aaai/text}. Finally, $\mathcal{R}$ is re-weighted by the output of convolution followed by a sigmoid activation for multi-level information fusing:
\begin{equation}
	\mathcal{R}^{'}=\mathcal{R} \odot \mbox{sigmoid}(f(\mathcal{R})),
\end{equation}
where $f(\cdot)$ is a learned convolution kernel with the size of $3\times 3$ in this work. After obtaining the multi-level relation matrix $\mathcal{R}^{'}$, the response of each similarity representation is updated according to its neighbors that are fused by hierarchical information. Besides, a residual connection is introduced to prevent gradient vanish. Therefore, the image-to-text level global alignment $\mathcal{S}^{*}_{i2t}$ can be obtained by
\begin{equation}
\mathcal{S}^{*}_{i2t}=W_r(\mathcal{R}^{'} \mathcal{S} W_g)+\mathcal{S},
\end{equation}
where $W_r$ and $W_g$ are learnable weight matrices. The number of HSR layers is set as $M$ $(M>=1)$, and the output of the previous layer is used as the input of the next layer to iteratively reason hierarchical similarity. The output of the global alignment at the last HSR layer is used as the final image-to-text similarity representation.

Regarding text-to-image level global alignment on the other hand, $s^{t2i}$ and $s^g$ are firstly concatenated to form an initial global feature $S^{t2i}=\{S_1^{t2i},\cdots,S_K^{t2i},S_{K+1}^{t2i}\}$. Different from image-to-text global alignment, a simple global average pooling operation is applied on $S^{t2i}$ to generate the text-to-image level global alignment $\mathcal{S}^{*}_{t2i}$ as
\begin{equation}
\mathcal{S}^{*}_{t2i}=\frac{1}{K+1}\sum_{j=1}^{K+1}{S_j^{t2i}}.
\end{equation}

At last, an element-wise addition operation is performed on $\mathcal{S}^{*}_{i2t}$ and $\mathcal{S}^{*}_{t2i}$ to integrate image-to-text and text-to-image level information to  generate the final global similarity representation $\mathcal{S}^{*}$ as
\begin{equation}
 \mathcal{S}^{*}=\mathcal{S}^{*}_{i2t} \oplus \mathcal{S}^{*}_{t2i}.
\end{equation}

\subsection{Loss Calculation}
\label{sect:prop:loss}

Following~\cite{2018arxiv/vsepp}, the bidirectional ranking loss is utilized to train the proposed TSHSR. A matched image-text pair $(v;t)$ is given within a mini-batch, the corresponding hardest negative image is denoted as $v^{-}$, and the hardest negative text is $t^{-}$. The bidirectional ranking loss is computed as
\begin{equation}
L\!=\![{\alpha}\!-\!S_\gamma{(v,t)}+S_\gamma{(v,t^{-})}]_{+} +[{\alpha}\!-\!S_\gamma{(v,t)}+S_\gamma{(v^{-},t)}]_{+},
\end{equation}
where $\alpha$ serves as a margin parameter, $[x]_{+}=\max(x,0)$, $S_\gamma{(\cdot)}$ is the similarity function in the joint embedding space, \emph{e.g.}, inner product used in this work.

\section{Experiment}
\label{sect:exp}

To verify the effectiveness of the proposed TSHSR approach, a large number of experiments are carried out on two benchmark datasets including MSCOCO~\cite{02eccv/mscoco} and Flickr30K~\cite{01tacl/flickr30k}. Section~\ref{sect:exp:data} introduces the datasets, evaluation metrics and implementation details. After that, Section~\ref{sect:exp:cmp} compares the proposed TSHSR with state-of-the-art methods. Furthermore, ablation study is performed to further assess the configurations of TSHSR in Section~\ref{sect:exp:absty}. Finally, several visualization results are exhibited in Section~\ref{sect:exp:vis}.

\subsection{Dataset and Setting}
\label{sect:exp:data}

\subsubsection{Dataset}
\label{sect:exp:data:ds}

The MSCOCO dataset~\cite{02eccv/mscoco} includes 123,287 images, and each image is annotated with 5 annotated captions. In this work, 113,287 images are chosen as the training set, 5000 images as the validation set and 5000 images as the testing set. The results are tested either by averaging over 5 folds of 1K test images or on the full 5K images. The Flickr30K dataset~\cite{01tacl/flickr30k} contains 31,783 images, and each image is also annotated with 5 captions. Following~\cite{02nips/DeViSE}, 29,783 images are split as the training set, 1000 images as the validation set and the rest as the testing set.

\subsubsection{Evaluation Metric}
\label{sect:exp:data:mt}

In this work, the image-text matching performance is measured by recall at $K$ (R@$K$), where the fraction of queries in which the correct item is retrieved in the closest $K$ points to the query.

\subsubsection{Implementation Detail}
\label{sect:exp:data:mt}

For image, the Faster-RCNN~\cite{03nips/faster-rcnn} detector is employed with ResNet-101~\cite{01cvpr/bottom-up} to extract the top $K=36$ salient regions, and each region proposal is encoded to a 2048-dimensional vector. For text, the word embedding size is set as 300, and the size of hidden states is 1024. The dimension of similarity representation is set as 256, with the temperature parameter ${\lambda}=9$, the number of hierarchical similarity reasoning steps $M=3$, and the margin parameter ${\alpha}=0.2$. The Adam optimizer~\cite{05iclr/adam} is employed to train TSHSR with the mini-batch size of 128 in an end-to-end manner. 20 epochs are trained on MSCOCO, the learning rate is set to be 0.0002 for the first 10 epochs initially and decays by 0.1 for the next 10 epochs. As for Flickr30K, 40 epochs are trained. For the first 30 epochs, the initial learning rate is also 0.0002, and the learning rate is decayed by 0.1 in the last 10 epochs. Snapshot is selected with the best
performance used by the summation of the recalls on the validation set.

\subsection{Comparison with State-of-the-Art Methods}
\label{sect:exp:cmp}

\begin{table*}[htbp]
    \renewcommand\arraystretch{1.1}
    \centering
    \caption{Comparison with state-of-the-art methods on MSCOCO 1K and Flickr30K. Best performance is indicated by the bold and runner-up by the underline.}
    \label{Table1}
    \newsavebox{\tableboxTabOV}
    \begin{lrbox}{\tableboxTabOV}
    \begin{tabular}{c|c|c|c|c|c|c|c|c|c|c|c|c}
        \hline
        \multirow{3}{*}{Methods} & \multicolumn{6}{c|}{MSCOCO 1K} & \multicolumn{6}{c}{Flick30K} \\
        \cline{2-13}
        & \multicolumn{3}{c|}{Sentence Retrieval} & \multicolumn{3}{c|}{Image Retrieval}
        & \multicolumn{3}{c|}{Sentence Retrieval} & \multicolumn{3}{c}{Image Retrieval} \\
        \cline{2-13}
        & R@1 & R@5 & R@10 & R@1 & R@5 & R@10 & R@1 & R@5 & R@10 & R@1 & R@5 & R@10 \\
        \hline
        CAMP\cite{01iccv/camp} & 72.3 & 94.8 & 98.3 & 58.5 & 87.9 & 95.0 & 68.1 & 89.7 & 95.2 & 51.5 & 77.1 & 85.3 \\
        SCAN\cite{01eccv/scan} & 72.7   & 94.8 & 98.4 & 58.8 & 88.4 & 94.8 & 67.4 & 90.3 & 95.8 & 48.6 & 77.7 & 85.2 \\
        SGM\cite{01wacv/csgm}   & 73.4  & 93.8 & 97.8 & 57.5 & 87.3 & 94.3 & 71.8 & 91.7 & 95.5 & 53.5 & 79.6 & 86.5 \\
        RDAN\cite{02ijcai/mvsa} & 74.6  & \underline{96.2} & \underline{98.7} & 61.6 & 89.2 & 94.7 & 68.1 & 91.0 & 95.9 & 54.1 & 80.9 & 87.2 \\
        MMCA\cite{14cvpr/Multi-Modality}& 74.8  & 95.6 & 97.7 & 61.6 & \underline{89.8} & 95.2 & 74.2 & 92.8 & 96.4 & 54.8 & \underline{81.4} & 87.8 \\
        BFAN\cite{02mm/BFAN} & 74.9 & 95.2 & - & 59.4 & 88.4 & - & 68.1 & 91.4 & - & 50.8 & 78.4 & - \\
        CAAN\cite{15cvpr/CAAN} & 75.5   & 95.4 & 98.5 & 61.3 & 89.7 & 95.2 & 70.1 & 91.6 & \textbf{97.2} & 52.8 & 79.0 & \underline{87.9} \\
        DPRNN\cite{02aaai/Expressing} & 75.3 & 95.8 & 98.6 & 62.5 & 89.7 & 95.1 & 70.2 & 91.6 & 95.8 & 55.5 & 81.3 & \textbf{88.2} \\
        PFAN\cite{03ijcai/pfan} & 76.5  & \textbf{96.3} & \textbf{99.0} & 61.6 & 89.6    & 95.2 & 70.0 & 91.8 & 95.0 & 50.4 & 78.7 & 86.1 \\
        VSRN\cite{02iccv/vsrn} & 76.2   & 94.8 & 98.2 & \underline{62.8} & 89.7 & 95.1 & 71.3 & 90.6 & 96.0 & 54.7 & \textbf{81.8} & \textbf{88.2} \\
        IMRAM\cite{09cvpr/imram} & 76.7 & 95.6  & 98.5 & 61.7 & 89.1 & 95.0 & 74.1 & \underline{93.0} & \underline{96.6} & 53.9 & 79.4 & 87.2 \\
        SGR\cite{01aaai/sgraf} & 78.0 & 95.8 & 98.2  & 61.4 & 89.3 & \underline{95.4} & 75.2 & \textbf{93.3} & \underline{96.6} & \underline{56.2} & 81.0 & 86.5 \\
        \hline
        Ours~(HSR) & \underline{78.8} & \textbf{96.3}    & 98.6 & \underline{62.8} & \textbf{89.9} & \textbf{95.6}    & \underline{76.0} & \underline{93.0} & 96.3 & 55.7 & 81.3 & 86.3 \\
        Ours~(TSHSR) & \textbf{79.0} & \underline{96.2} & 98.6 & \textbf{63.1} & \textbf{89.9}   & \underline{95.4} & \textbf{76.3} & \underline{93.0} & 95.8    & \textbf{56.6} & 81.2 & 85.9 \\
        \hline
    \end{tabular}
    \end{lrbox}
    \scalebox{1.0}{\usebox{\tableboxTabOV}}
\end{table*}

\subsubsection{Results on MSCOCO 1K and Flickr30K}
\label{sect:exp:cmp:p1}

Table~\ref{Table1} shows the quantitative results on MSCOCO 1K and Flickr30K, where it can be seen that the proposed TSHSR outperforms all of the state-of-the-art methods with a large gap on R@1. Concretely, TSHSR has the best R@1=79.0\% for sentence retrieval and R@1=63.1\% for image retrieval on MSCOCO 1K. As for Flickr30K, TSHSR also achieves the best on R@1, with the best R@1=76.3\% for sentence retrieval and R@1=56.6\% for image retrieval. In addition, when the HSR module is used alone for image-to-text level similarity representation, namely ``Ours~(HSR)'' in Table~\ref{Table1}, it also performs well, with R@1=78.8\% for sentence retrieval and R@1=62.8\% for image retrieval on MSCOCO 1K, R@1=76.0\% for sentence retrieval and R@1=55.7\% for image retrieval on Flickr30K. According to the results, two conclusions can be obtained. First, HSR improves the capability of image-text matching, which verifies the importance of utilizing hierarchical interaction patterns. Second, when image-to-text level and text-to-image level similarity representations are both considered, namely ``Ours~(TSHSR)'' in Table~\ref{Table1}, it not only improves the performance of image retrieval, but also boosts the performance of sentence retrieval, which indicates that image-to-text level and text-to-image level similarity representations are complementary.

\subsubsection{Results on MSCOCO 5K}
\label{sect:exp:cmp:p2}

Table~\ref{Table2} shows the quantitative results on MSCOCO 5K. It is noteworthy that TSHSR also outperforms all of the state-of-the-art methods with about 1\% improvement on R@1. Besides, HSR also obtains a competitive retrieval performance among almost all of the state-of-the-art methods, demonstrating the necessity and effectiveness of hierarchical similarity reasoning.

\begin{table}[htbp]
    \renewcommand\arraystretch{1.1}
    \centering
    \caption{Comparison with state-of-the-art methods on MSCOCO 5K. Best performance is indicated by the bold and runner-up by the underline.}
    \label{Table2}
    \begin{lrbox}{\tableboxTabOV}
    \begin{tabular}{c|c|c|c|c}
        \hline
        \multirow{3}{*}{Methods} & \multicolumn{4}{c}{MSCOCO 5K} \\
        \cline{2-5}
        & \multicolumn{2}{c|}{Sentence Retrieval} & \multicolumn{2}{c}{Image Retrieval} \\
        \cline{2-5}
        & R@1 & R@10 & R@1 & R@10 \\
        \hline
        SGM\cite{01wacv/csgm}   & 50.0 & 87.9 & 35.3 & 76.5 \\
        CAMP\cite{01iccv/camp} & 50.1   & 89.7 & 39.0 & 80.2 \\
        SCAN\cite{01eccv/scan} & 50.4 & 90.0 & 38.6 & 80.4 \\
        CAAN\cite{15cvpr/CAAN} & 52.5   & 90.9 & \underline{41.2} & \textbf{82.9} \\
        VSRN\cite{02iccv/vsrn} & 53.0   & 89.4 & 40.5 & 81.1 \\
        IMRAM\cite{09cvpr/imram} & 53.7 & \underline{91.0} & 39.7 & 79.8 \\
        MMCA\cite{14cvpr/Multi-Modality} & 54.0 & 90.7 & 38.7 & 80.8 \\
        SGR\cite{01aaai/sgraf}   & \underline{56.9} & 90.5 & 40.2 & 79.8 \\
        \hline
        Ours~(HSR) & 56.6 & \textbf{91.4} & 40.8 & 80.9 \\
        Ours~(TSHGR) & \textbf{57.4} & \textbf{91.4} & \textbf{41.4} & \underline{81.2} \\
        \hline
    \end{tabular}
    \end{lrbox}
    \scalebox{1.0}{\usebox{\tableboxTabOV}}
\end{table}

\subsection{Ablation Study}
\label{sect:exp:absty}

To further explore the impact of different configurations about the proposed TSHSR, ablation studies are conducted on the MSCOCO 1K dataset.

\begin{table}[htbp]
    \renewcommand\arraystretch{1.1}
    \centering
    \caption{Comparison with different number of HSR layers.}
    \label{Table3}
    \begin{lrbox}{\tableboxTabOV}
    \begin{tabular}{c|c|c|c|c|c|c|c|c}
        \hline
        \multirow{2}{*}{Model} & \multicolumn{4}{c|}{Number} & \multicolumn{2}{c|}{Sentence Retrieval}
        & \multicolumn{2}{c}{Image Retrieval} \\
        \cline{2-9}
        & 0 & 1 & 2 & 3 & R@1 & R@10 & R@1 & R@10 \\
        \hline
        1 & \checkmark &  &  &  & 77.8 & 98.5 & 62.0 & 95.3 \\
        \hline
        2 &  & \checkmark &  &  & 78.2  & 98.5 & 62.2 & 95.4 \\
        \hline
        3 &  &  & \checkmark &  & 78.7  & 98.5 & 62.6 & 95.5 \\
        \hline
        4 &  &  &  & \checkmark & \textbf{78.8} & \textbf{98.6}
        & \textbf{62.8} & \textbf{95.6} \\
        \hline
    \end{tabular}
    \end{lrbox}
    \scalebox{1.0}{\usebox{\tableboxTabOV}}
\end{table}

\subsubsection{How effective is the number of reasoning steps using HSR}

To explore the impact of the number of HSR layers, four numbers are tried as listed in Table~\ref{Table3}. From the results, it can be seen that with the increase of HSR layers, the performances of both sentence retrieval and image retrieval are improved gradually. This reveals that HSR benefits to both image retrieval and sentence retrieval.

\begin{table}[htbp]
    \renewcommand\arraystretch{1.1}
    \centering
    \caption{Comparison of different configurations of similarity reasoning. In the first row, `0' indicates traditional similarity reasoning, and `1' indicates the proposed hierarchical similarity reasoning.}
    \label{Table4}
    \begin{lrbox}{\tableboxTabOV}
    \begin{tabular}{c|c|c|c|c|c|c}
        \hline
        \multirow{2}{*}{\diagbox[width=9.5em]{Steps}{Config.}} & \multirow{2}{*}{0} & \multirow{2}{*}{1}
        & \multicolumn{2}{c|}{Sen. Ret.} & \multicolumn{2}{c}{Img. Ret.} \\
        \cline{4-7}
        &  &  & R@1 & R@10 & R@1 & R@10 \\
        \hline
        \multirow{2}{*}{1} & \checkmark &  & 78.1 & 98.5 & 62.4 & 95.4 \\
        \cline{2-7}
        &  & \checkmark & 78.2 & 98.5 & 62.2 & 95.4 \\
        \hline
        \multirow{2}{*}{3} & \checkmark &  & 78.3 & 98.3 & 62.8 & 95.5 \\
        \cline{2-7}
        &  & \checkmark & 78.8 & 98.6 & 62.8 & 95.6 \\
        \hline
    \end{tabular}
    \end{lrbox}
    \scalebox{1.0}{\usebox{\tableboxTabOV}}
\end{table}

\begin{table*}[htbp]
    \renewcommand\arraystretch{1.1}
    \centering
    \caption{Comparison with different level representations when using HSR. I2T indicates reasoning on image-to-text level similarity, and T2I indicates reasoning on text-to-image level similarity.}
    \label{Table5}
    \begin{lrbox}{\tableboxTabOV}
    \begin{tabular}{c|c|c|c|c|c|c|c|c}
        \hline
        \multirow{2}*{\diagbox{Alignment}{Steps}}
        & \multirow{2}*{1} & \multirow{2}*{2} & \multirow{2}*{3}
        & \multicolumn{2}{c|}{Sen. Ret.} & \multicolumn{2}{c|}{Img. Ret.} & \multirow{2}*{Sum} \\
        \cline{5-8}
        &  &  &  & R@1 & R@10 & R@1 & R@10 & \\
        \hline
        \multirow{3}{*}{I2T} & \checkmark &  &  & 78.2 & 98.5 & 62.2 & 95.4 & 334.3 \\
        \cline{2-9}
        &  & \checkmark &  & 78.7 & 98.5 & 62.6 & 95.5  & 335.3 \\
        \cline{2-9}
        &  &  & \checkmark & 78.8 & 98.6 & 62.8 & 95.6 & \textbf{335.8} \\
        \hline
        \multirow{3}{*}{T2I} & \checkmark &  &  & 78.1 & 98.6 & 61.6 & 95.2 & 333.5 \\
        \cline{2-9}
        &  & \checkmark &  & 77.7 & 98.2 & 61.5 & 95.3 & 332.7 \\
        \cline{2-9}
        &  &  & \checkmark & 79.0 & 98.4 & 62.0 & 95.4 & 334.8 \\
        \hline
    \end{tabular}
    \end{lrbox}
    \scalebox{1.0}{\usebox{\tableboxTabOV}}
\end{table*}

\begin{figure*}[htbp]
    \centering
    \includegraphics[width=0.9\textwidth]{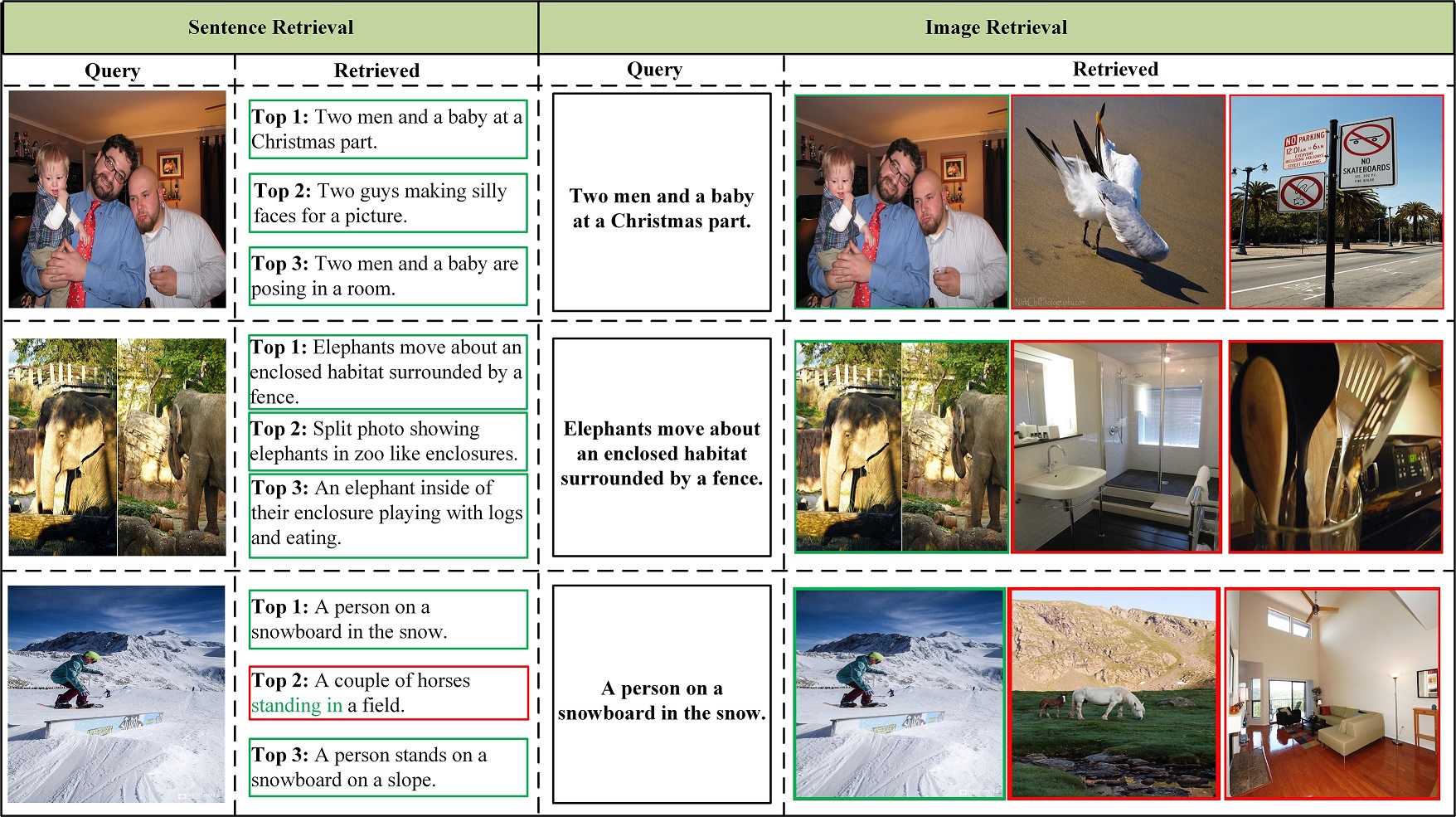}
    \caption{Visualization of image-text mutual retrieval by TSHSR on MSCOCO. The top-3 ranked text captions are shown for each image query, and the top-3 ranked images are also illustrated for given text query. Correct matched samples are framed by green boxes while incorrect matched samples are framed by red boxes.}
    \label{visualization}
\end{figure*}

\subsubsection{How necessary is hierarchical information}

To explore how important is the hierarchical information for reasoning, several experiments are designed by  comparing the performance between hierarchical similarity reasoning and traditional similarity reasoning with different number of reasoning steps. The results are shown in Table~\ref{Table4}, where it can be observed that the performances of using hierarchical information are superior to that without using, demonstrating the necessity of hierarchical information for reasoning.

\subsubsection{How important are different level representations}

To explore the importance of different level similarity contributed to image-text matching, an ablation study is conducted with the comparative results shown in Table~\ref{Table5}. The results from Table~\ref{Table5} show that reasoning on image-to-text level similarity can obtain better performance than text-to-image level similarity. This is possibly due to the reason that image-to-text level similarity is aligned with natural language information and text-to-image level similarity is aligned with visual information, while natural language information is simpler than visual information.  More important, image-to-text level similarity still retains structured information (such as sequential information), which can be extracted easily by convolutional operation. However, the structured information (such as spatial information) about text-to-image level similarity is destroyed.

\subsection{Visualization of Retrieved Results}
\label{sect:exp:vis}

As shown in Fig.~\ref{visualization}, to validate the mutual retrieval performance achieved by TSHSR, the top-3 ranked items are listed according to image query and text query. Specifically, it can be easily found that the top-1 retrieved results are all right not only in sentence retrieval but also in image retrieval, which verifies the superiority of the proposed TSHSR. We also show some incorrect retrieved results in Fig.~\ref{visualization}. For example, the top-2 ranked item in the third sentence retrieval is ``A couple of horses standing in the field'', which misunderstands the person as a horse. More importantly, TSHSR has the ability to capture complicated matching patterns. For instance, in the first sample of sentence retrieval, TSHSR not only captures the entities of three persons as shown in the top-1 ranked item, but also detects the tiny expression as shown in the top-2 ranked item.

\section{Conclusion}
\label{sect:cln}

In order to tackle the two problems of insufficient hierarchical information and inadequate information utilization in the image-text matching task, a simple and effective architecture TSHSR is presented to enable the reasoning procedure to spontaneously employ hierarchical information and utilize both image-to-text level as well as text-to-image level similarity information. The proposed TSHSR is evaluated on two benchmark datasets, and the comparative results demonstrate the effectiveness of the design principle and the better performance over other state-of-the-art methods. Although current reasoning approaches can construct the relationship between two entities, such as ``move'', ``play'', ``show'', etc, positional information is usually overlooked in the reasoning process, such as ``in'', ``on'', ``over'' and so on. In the future, it is desired to explore the importance of positional information for reasoning.

\bibliographystyle{IEEEtran}
\bibliography{ref}

\end{document}